\documentclass[aps,prb,twocolumn,showpacs,amsmath,amssymb,floatfix]{revtex4}
\usepackage{graphicx}
\usepackage{bm}
\def\be{\begin{equation}}
\def\ee{\end{equation}}
\def\bea{\begin{eqnarray}}
\def\eea{\end{eqnarray}}
\def\la{\langle}
\def\ra{\rangle}

\begin{document}
\title{Weak measurement of quantum dot spin qubits}
\author{Andrew N. Jordan$^1$, Bj\"orn Trauzettel$^2$, and Guido Burkard$^{2,3}$}
\affiliation{
$^1$\ Department of Physics and Astronomy, University of Rochester, Rochester, New York 14627, USA}
\affiliation{
$^2$\ Department of Physics and Astronomy, University of Basel,
Klingelbergstrasse 82, CH-4056 Basel, Switzerland}
\affiliation{
$^3$\ Institute of Theoretical Physics C, RWTH Aachen University,
D-52056 Aachen, Germany}

\date{June 1, 2007}

\begin{abstract}
The theory of weak quantum measurements is developed for quantum dot spin
qubits.  Building on recent experiments, we propose a control cycle to
prepare, manipulate, weakly measure, and perform quantum state tomography.  This is accomplished 
using a combination of the physics of electron spin resonance, spin blockade,
and Coulomb blockade, resulting in a charge transport process.  We investigate
the influence of the surrounding nuclear spin environment, and find a regime
where this environment significantly simplifies the dynamics of the weak
measurement process, making this theoretical proposal realistic with existing
experimental technology.  We further consider spin-echo refocusing to combat
dephasing, as well as discuss a realization of ``quantum undemolition'',
whereby the effects of quantum state disturbance are undone. 
\end{abstract}
\maketitle

\section{Introduction}

Continuous weak measurement has attracted great interest recently, not only
because the phenomenon sheds light on fundamental physics, but also for its
possible application to practical tasks in computation, state preparation, and
error correction.  While the informational theory of weak measurement has been
under active development in quantum dot charge qubits,\cite{charge1,charge2,charge3} the
theory has not been 
well developed for spin qubits, a major area of experimental activity. The
purpose of this paper is to develop the theory of weak measurements for spin
qubits, both regarding the manner in which the state is affected by weak
measurement and new applications that can be developed with controlled weak
quantum measurements.  Importantly, the modern theory of weak measurements has
recently been experimentally verified in the solid state by the Martinis group.  This experiment 
investigated weak measurements in superconducting phase qubits by utilizing quantum state tomography
of the post-measurement state.\cite{Katz}   

The use of electron spins as quantum bits is very attractive in view of their
ability to be effectively isolated from the environment for relatively long
times.\cite{loss}  These long coherence times are due in part to the small magnetic moment of the electron.  A small magnetic moment also poses a problem for single spin read-out.  This was overcome by the use of spin-to-charge conversion;\cite{elzerman,hanson} a technique\cite{loss} where the spin information is first converted into charge information which is subsequently detected, using {\it e.g.} a quantum point contact.
A second major problem is how to couple two nearby spins, considering the very weak direct magnetic dipole interaction. 
This difficulty was overcome by using the charge-mediated exchange coupling.\cite{petta}  The latest experimental accomplishment demonstrates single-spin manipulation with (magnetic) electron spin resonance (ESR).\cite{leo}  In this experiment, it was shown that short bursts of oscillating magnetic field can drive coherent Rabi oscillations in the individual electron spins confined to a quantum dot.

All the ingredients for universal quantum computation are now available in
this system.   However, there has been recent theoretical activity indicating
that there may be significant practical advantages to using weak continuous measurement 
over projective measurements.  For example, it has been shown that rather than
using two-qubit unitary operations plus projective single-qubit measurements,
that a two-qubit parity meter\cite{parity1,parity2,myparity1,myparity2} (where only the parity
subspace of the 2-qubit Hilbert space is able to be resolved) plus
single-qubit measurements is sufficient to enable universal quantum
computation\cite{parity1,parity2,myparity1,myparity2}
as well as create fully entangled Bell states.\cite{myparity1,myparity2,ruskov1,mao} 
This discovery eliminates the need for 2-qubit
unitaries thus avoiding the necessity of strong (direct) qubit-qubit interactions.
This one example is sufficient impetus to justify intensive investigation into
weak measurements for spin qubits.  


\begin{figure*}[htb]
\begin{center}
\leavevmode \includegraphics[width=14cm]{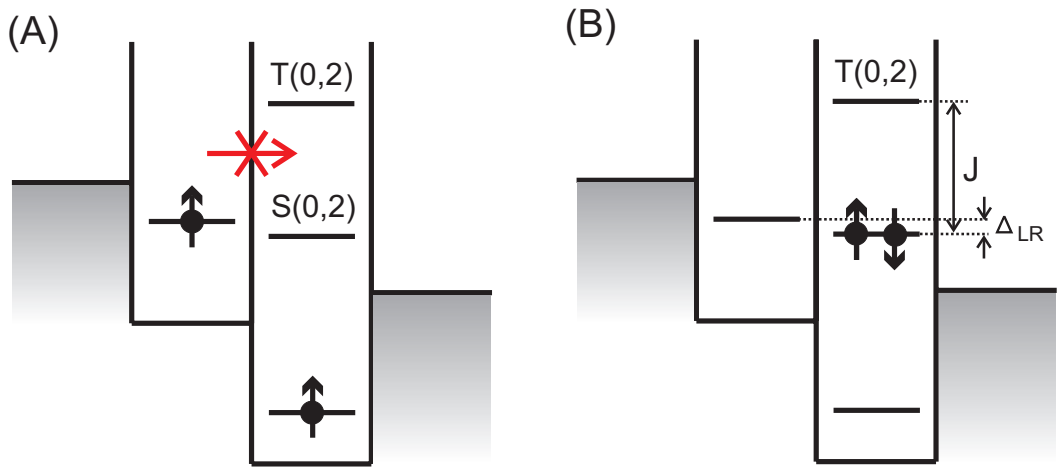}
\caption{(Color online) The figure illustrates the two possible two-electron double dot
  states that appear in the weak measurement setup (as described in the
  text). {\bf (A)} The double dot is in the $(1,1)$ configuration and the two
  electron spins form a triplet. Then, the electron in the left dot cannot
  tunnel into the right dot to form a $(0,2)$ state because the triplet state
  $T(0,2)$ of the $(0,2)$ configuration is energetically too high, putting it 
  outside the transport energy window. The energy
  difference between $T(0,2)$ and $S(0,2)$ is the spin exchange coupling $J$. The
  energy difference $\Delta_{\rm LR}$ between the (1,1) and the
  $S(0,2)$ states (which is chosen to be much smaller than $eV$ in the figure, where
  $V$ is the applied bias) can be tuned by external gates that shift the
  energy levels of each dot independently. Therefore, $\Delta_{\rm LR}$ can,
  in principle, take any desired value.
  {\bf (B)} If the
  two electron spins in the (1,1) configuration form a singlet $S(0,2)$, then
  tunneling from the (1,1) to the (0,2) configuration is energetically allowed
  and the resulting state is shown. It is possible to unblock the state shown
  in {\bf (A)}
  and to allow for the transition to the state shown in {\bf (B)} 
  by applying an electron spin resonance (ESR) signal to
  the left dot while the two electrons are in the spin-blocked state
  $T(1,1)$.} \label{fig1}
\end{center}
\vspace{-5mm}
\end{figure*}

We will now describe some of the details of the recent experiment of Koppens {\it et al.} that we propose
to extend.\cite{leo}  The qubit is encoded with two electron spins, where each electron is confined in a separate quantum dot.  Electrical bias is applied across the double quantum dot,
where the right dot is lowered energetically below the left 
dot with a gate voltage.   The notation $(n, m)$ refers to $n$ electrons
occupying the left dot, and $m$ electrons occupying the right dot.
Electrons can tunnel from the left lead to the left dot with a rate
$\Gamma_L$.  The dots are tuned to the Coulomb blockade (CB) regime such that only
the states $(0,1)$, $(1,1)$, and $(0,2)$ can be occupied during a transport
cycle. The gate voltages applied to the quantum dot structure are tuned such that the sequential tunneling cycle $(0,1)\rightarrow(1,1)\rightarrow(0,2)\rightarrow(0,1)$ is energetically allowed.   This cycle consists of a first step, where an electron hops onto the left quantum dot, a second step where an electron hops from the left to the right dot (which has been occupied previously by a single electron) with rate $\Gamma$, 
and finally a third step which closes the cycle and in which one of the electrons on the right dot hops out into the right reservoir
with rate $\Gamma_R$.  In this sequential tunneling configuration, {\it spin blockade} further restricts transport to situations where the two electrons form a spin singlet (0,2)S on the right dot while the spin triplet (0,2)T is outside the transport energy window due to the large single-QD exchange energy $J$, see Fig.~1(B).
If the electrons are in any of the triplet states (1,1)T, then the current is blocked since the electron in the left dot can neither tunnel to the right nor to the left as illustrated in Fig.~1(A).
Once this spin blockade state is reached, the gate voltages are adjusted such
that the system is now in the Coulomb blockade (CB) regime, where sequential
transport is suppressed by the interaction between the electrons and the
occupation numbers on the dots are fixed to (1,1).  Then, an ESR pulse is used
to prepare a superposition of the singlet and triplet states (Fig.~2).  In the
CB regime, any unwanted tunneling events between the dots and the leads that
could lead to spin flips are suppressed.  After the ESR pulse, the system is
brought back into the sequential transport regime which now allows for a
coherent weak measurement of the prepared state.

This setup is naturally suited to investigate weak quantum measurements.  The
measurement scheme we now describe is closely related to recent developments
in superconducting phase qubits\cite{Martinis,Katz} where the readout process
also involves a quantum tunneling process.   The essential idea is to
introduce another time scale into the measurement process.  By waiting for a
time much longer than the average inter-dot tunneling time
$\Gamma^{-1}$, one projects the system with certainty 
into either the triplet subspace, or the singlet state.   However, if it is
possible to let the system ``try to tunnel'' for a time comparable to
$\Gamma^{-1}$ , then the measurement is weak.   We will give the details of
how this happens below. 

The physical process described above may be mathematically described by introducing a measurement operator $M_Q$ that describes the physical weak measurement experienced by the spin qubit, such that the probability of either event given an initial density matrix $\rho$ is given by 
\be
P(Q) = {\rm Tr} \rho M_Q^\dagger M_Q,
\ee
where $Q=0$ if no electron has tunneled and $Q=1$ if an electron has tunneled.  
Quantum mechanics then predicts that coherent, yet nonunitary evolution of the density matrix under the condition that measurement result $Q$ is found, is given by 
\be
\rho' = M_Q\rho M_Q^\dagger/P(Q),
\ee
(see {\it e.g.} Ref.~\onlinecite{book}) where the positive operator-valued measure (POVM) elements $E_Q = M_Q^\dagger M_Q$ must obey completeness, $\sum_Q E_Q =1$.  One of the main differences compared to the superconducting phase qubit example already demonstrated\cite{Katz} is the fact that the informational spin qubit is encoded into two microscopic electrons (two physical qubits), rather than just one macroscopic qubit.  Another difference is the fact that the spin readout mechanism is via a charge transport process, rather than a change of magnetic flux.

We anticipate that the development of the theory of weak quantum measurements for spin qubits will play a key role in future experimental investigations, as full quantum control is mastered.  

\section{Minimal Model}

We first consider the simplest case of no environmental decoherence from the surrounding nuclear spins, and no inelastic transitions.  We also assume for simplicity that $\Gamma \ll \Gamma_L, \Gamma_R$, so the central barrier is the bottle-neck in the transport cycle.
When both dots are occupied by one electron, $(1,1)$, we define the triplet $(T)$ and singlet $(S)$ states as
\bea
T_0 &=&  (|\!\uparrow \downarrow \rangle + |\!\downarrow \uparrow \rangle )/\sqrt{2}, \quad T_+ =  |\!\uparrow \uparrow \rangle , \quad T_- =  |\!\downarrow \downarrow \rangle,  \nonumber \\
S &=&  (|\!\uparrow \downarrow \rangle - |\!\downarrow \uparrow \rangle )/\sqrt{2}.
\eea

We now follow a modification of the control cycle described in Ref.~\onlinecite{leo} (see Fig.~2): 

({\it i}) We consider the initial state to be $T_+$ (which of the triplet states is chosen
is not really important), so the transport cycle is blocked.  

({\it ii}) Next, we lower the gate voltage on the left dot, putting the system
into Coulomb blockade (none of the levels in the right dot are accessible,
forbidding all transitions), and turn on the ESR signal that induces single
spin rotations on the left qubit (here we assume the ESR pulse is on resonance
with the left spin only, and make the rotating wave approximation).   In particular, in the rotating frame 
\be
|\!\uparrow \rangle_L \rightarrow  \cos \theta_1 |\!\uparrow  \rangle_L +\sin \theta_1 \vert\!\downarrow \rangle_L,
\ee
where $\theta_1=\Omega \tau$, $\Omega$ is the Rabi flopping frequency, and
$\tau$ is the time it is on for.  Therefore, after the ESR pulse, the two-spin
state is $|\psi_0\ra = \cos \theta_1   |\!\uparrow \uparrow \rangle + \sin
\theta_1 |\!\downarrow \uparrow \rangle  = \cos \theta_1\ T_+ + \sin \theta_1\
(T_0-S)/\sqrt{2} $.


\begin{figure*}[htb]
\begin{center}
\leavevmode \includegraphics[width=8cm]{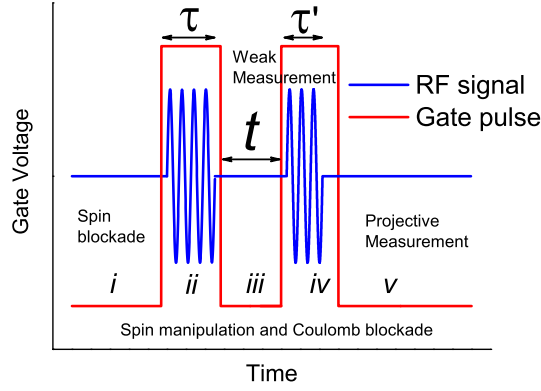} 
\caption{(Color online) Control cycle for the weak measurement demonstration. {\bf (i)} The system is brought into the spin blockade regime, see Fig.~\ref{fig1}
  {\bf (A)}. {\bf(ii)} A gate pulse is used to tune the system into
  Coulomb blockade and a RF signal is applied. This is the state preparation
  step. {\bf (iii)} In the weak measurement step, tunneling from the left to
  the right dot is possible if the system is in the singlet
  configuration. This step lasts shorter than $\Gamma^{-1}$. {\bf (iv)}  A gate pulse is used to tune the system into
  Coulomb blockade and a RF signal is applied. This is the state
  tomography step. {\bf (v)} In the projective measurement step, tunneling
  from the left to the right dot is possible if the system was in the singlet
  configuration. This step lasts longer than $\Gamma^{-1}$.} \label{fig2}
\end{center}
\vspace{-5mm}
\end{figure*}

({\it iii}) The next step is the raising of the left gate voltage, allowing the system to tunnel in a state-selective way.  After the left electron enters the right dot [forming the state $S(0,2)$], the much smaller right tunneling rate causes the escape of the electron to the right lead, leaving $(0,1)$. 
As mentioned before, the transitions $T_{0,-,+}(1, 1) \rightarrow S(0, 2)$ are forbidden because the levels in the right dot for these transitions are energetically inaccessible (giving no transported charge, $Q=0$), but the transition $S(1, 1) \rightarrow S(0, 2)$ is allowed (giving a transported charge, $Q=1$) with rate $\Gamma$.  Because tunneling is an exponential decay process, we can express this mathematically by saying that in the singlet-triplet basis $(T_+, T_-, T_0, S)$, the singlet-singlet matrix element of the POVM element is
\be
\la S \vert M^\dagger_Q M_Q \vert S\ra = \begin{cases} \exp(-\Gamma t), & Q=0 \\ 1- \exp(-\Gamma t), & Q=1 \end{cases}, \label{s-POVM}
\ee
while the triplet-triplet matrix elements are
\be
\la T_j \vert M^\dagger_Q M_Q \vert T_j \ra = \begin{cases} 1, & Q=0 \\ 0, & Q=1 \end{cases},
\ee
where $j=+,-,0$.  The off-diagonal matrix elements vanish in this basis.  
Considering a pure tunneling process that does not induce any phase, we can write the measurement operators in the singlet/triplet basis simply as the square root of the POVM element.

Therefore, the post-measurement state of the qubit is (for pure states) $\psi'_Q = M_Q \psi_i /{\cal N}$, where ${\cal N}$ is the (re-)normalization of the new state, so that
$P_{Q=0} = |{\cal N}|^2$.  If $Q=1$, the state is destroyed (the configuration is now $(0,2)$ which will quickly be followed by the electron tunneling and going to the drain). 
The probability of finding an electron in the drain $(Q=1)$ at this step is then
\be
P_{iii}(1) = \la \psi_0 \vert M^\dagger_{1} M_{1} \vert \psi_0 \ra = \sin^2\theta_1 [1-\exp(-\Gamma t)]/2.
\ee
The probability of not finding an electron in the drain $(Q=0)$ at this step is 
\bea
P_{iii}(0) &=& \la \psi_0 \vert M^\dagger_{0} M_{0} \vert \psi_0 \ra \nonumber \\
&=& \cos^2 \theta_1 + \sin^2 \theta_1 [1+\exp(-\Gamma t)]/2.
\eea
Notice that $P_{iii}(0)+P_{iii}(1)=\cos^2 \theta_1 + \sin^2 \theta_1 = 1$.
In the null-result ($Q=0$) case, the post-measurement state is 
\be
\psi_0' =  \frac{1}{\sqrt{2}{\cal N}}\begin{pmatrix}\sqrt{2} \cos \theta_1 \\ 0 \\ \sin \theta_1 \\ -\sin \theta_1 \exp(-\Gamma t/2) \end{pmatrix},
\label{post}
\ee
where ${\cal N}^2 =\cos^2 \theta_1 +\sin^2 \theta_1 D_+$, and we define
\be
D_{\pm} = [1 \pm \exp(-\Gamma t/2)]/2.
\ee

  If no time has elapsed, then the new state is identical to the initial
  state, while in the long time limit, $\Gamma t \gg 1$, the singlet portion
  of the state is continuously removed. Qualitatively, this is because if no
  charge is seen to be transfered after a sufficiently long time, we can be
  confident that the quantum state must be somewhere in the triplet subspace,
  but we gain no information about which triplet state the system is in. 

({\it iv})  In order to confirm that the state (\ref{post}) is indeed the
post-measurement state, we can apply quantum state tomography by first
applying another ESR pulse, and then a second (projective) measurement.\cite{tomo}  
Because the ESR pulse acts in the left/right basis, and not the
singlet/triplet basis, it is first necessary to return to the left/right
basis.  The unitary operation $U_2$ that converts the basis $(T_+, T_-, T_0,
S)$ to $(\vert\!\uparrow \uparrow\ra, |\!\downarrow \downarrow\ra, |\!\uparrow
\downarrow\ra, |\!\downarrow \uparrow\ra)$ is 
\be
U_2 = \begin{pmatrix}  1 & 0 & 0 & 0 \\ 0 & 1 & 0 & 0 \\0 & 0 & \frac{1}{\sqrt{2}} & \frac{1}{\sqrt{2}} \\
0 & 0 & \frac{1}{\sqrt{2}} & - \frac{1}{\sqrt{2}}  \end{pmatrix}.
\ee
Applying this matrix to the state (\ref{post}), we find that in the left/right basis
\be
\psi'_{L/R} =  \frac{1}{\cal N}\begin{pmatrix} \cos \theta_1 \\ 0 \\ \sin \theta_1 D_- 
\\ \sin \theta_1 D_+ \end{pmatrix}.
\label{postlr}
\ee
We can now implement the ESR-pulse on the left spin by applying the $SU(2)$ rotation matrix 
\be
R_L = \begin{pmatrix}  \cos \theta_2 & -\sin \theta_2 \\ \sin \theta_2 & \cos \theta_2 \end{pmatrix}
\label{Rl}
\ee
to the left spin only, where $\theta_2=\Omega \tau'$ represents the angle the left spin is driven through in the rotating frame.  This produces the state
\be
\psi_{L/R}^{\rm final} =  \frac{1}{\cal N}\begin{pmatrix} \cos \theta_1 \cos \theta_2 - \sin \theta_1 \sin \theta_2 D_+  \\  \sin \theta_1 \sin \theta_2 D_-   \\ \sin \theta_1 \cos \theta_2 D_- \\ \cos \theta_1 \sin \theta_2 + \sin \theta_1 \cos \theta_2 D_+ \end{pmatrix}.
\label{final}
\ee

({\it v}) Make a projective measurement:  Now we lower the gate voltage again, allowing the left electron to tunnel to the right well (and escape to the drain), this time keeping the voltage low for a time much longer than the inverse tunneling rate.
The probability that the tunneling event will occur is given by the square
overlap between the state (\ref{final}) and the singlet state, $P_v(1) = \vert
\la S \vert \psi_{L/R}^{\rm final} \ra\vert^2$.  The final state in the
triplet/singlet basis is given by applying the inverse of $U_2$, so we find
for the probability $P_v(1)$ of tunneling in the second (strong) measurement: 
\be
P_v(1) = \frac{\left[ \cos \theta_1 \sin \theta_2 + \sin \theta_1 \cos\theta_2 \exp(-\Gamma t/2)\right]^2}{2 P_{iii}(0)},
\label{pfinal}
\ee
where we recall $P_{iii}(0)={\cal N}^2= \cos^2 \theta_1 +  \sin^2\theta_1 D_+$.

We are now in a position to compute the total probability of finding a transported electron through the whole cycle.
This is given by the probability the tunneling event occurred in step $(iii)$ or the probability the tunneling event did not occur in step $(iii)$, but did occur in step $(v)$.  Therefore the total probability is given by
\bea
P_{\rm tot} &=& P_{iii}(1) + P_{iii}(0) P_v(1)\\ \label{tot}
&=&  \sin^2\theta_1  [1-\exp(-\Gamma t)]/2 \nonumber \\ &+& [\cos \theta_1 \sin \theta_2 + \sin \theta_1 \cos \theta_2 \exp(-\Gamma t/2)]^2/2. \nonumber
\eea
This result is naturally interpreted in terms of a state preparation step, characterized by $\theta_1$, the weak measurement, characterized by a strength $\Gamma t$, and a tomography step, characterized by an angle $\theta_2$.

This analysis describes one cycle.  The experiment is now repeated many times, with a cycle period $T$, and the average current is measured at fixed weak measurement times, and rotation angles. The average current is given by the total probability of a successful tunneling event, divided by the cycle time,
\be
\la I \ra = \frac{e P_{\rm tot}}{T}.
\ee
Here we see another attractive feature of the proposal:  there is no need for
statistical averaging over large data sets as in Ref.~\onlinecite{Katz}; the system
self-averages and gives the final answer (\ref{tot}) as a small electrical
current.  

Generalizing to the situation where the initial state is in any coherent
superposition $\psi_0 = \alpha T_0 + \beta T_+ + \gamma T_-$ of triplet states
and repeating the previous steps, we find the total probability is given by
that same result, Eq.~(\ref{tot}), but weighted by the overall factor $\vert
\beta +\gamma\vert^2$.  This indicates that if the initial state was $T_0$, no
electrons could be transfered with this sequence.

There is the possibility that between cycles there can be the uninterrupted
cycle: $(0,1) \rightarrow S(1,1)\rightarrow S(0,2) \rightarrow (0,1)$.  This
will contribute a background current to the signal that must be subtracted.  We
note that the choice $\theta_1 = \theta_2 = \pi$ gives a vanishing signal for
all weak measurement times $t$, providing a calibration point. 

In the transport cycle, the system will get blocked in statistically
independent triplet states, and therefore we should average the total
tunneling probability over an ensemble of initial triplet states.  Taking the
average with a completely mixed density matrix indicates that
$\la|\alpha|^2\ra_{r} = \la|\beta|^2\ra_r= \la |\gamma|^2\ra_r=1/3$, where
$\la\ldots\ra_r$ denotes averaging over statistically independent
realizations, while the coherences average to zero in repeated realizations.
Applying this average to the general probability, we find again the result
(\ref{tot}), but multiplied by $2/3$. 

\section{Influence of the inhomogeneous nuclear magnetic field}

We now turn to a more realistic treatment of the physics by including the
effect of the surrounding environment. The dominant source of dephasing in
GaAs quantum dot spin qubit is interaction with the surrounding nuclear spins.
This has been theoretically analyzed in detail in the past few years
\cite{nuclear_theo1,nuclear_theo2,nuclear_theo3,nuclear_theo4} 
and also experimentally verified.\cite{nuclear_exp1,nuclear_exp2}
The dynamics of the nuclear system is much slower than the electron spin
dynamics, so in a given run the composite nuclear magnetic field is
essentially static.  This gives rise to a systematic (unknown) unitary
rotation which is taken into account below.  The magnetic field changes in
different realizations of the measurement cycle, leading to an effective
dephasing when the data is averaged over a statistical ensemble. We will
discuss dephasing in more detail below. 

To take these effects into account, the Hamiltonian of the spins interacting with the external ${\bf B}_{\rm ext}$, nuclear, and oscillating ${\bf B}_{\rm ac}$ magnetic field is
\bea
H &=& g \mu_B ({\bf B}_{\rm ext} + {\bf B}_{L,N}) {\bf S}_L  + g \mu_B ({\bf B}_{\rm ext} 
+ {\bf B}_{R,N}) {\bf S}_R \nonumber \\ &+& g \mu_B \cos (\Omega t) {\bf B}_{\rm ac} (g_L {\bf S}_L + g_R {\bf S}_R).\label{H-nucl}
\eea
Here ${\bf B}_{L,N}$ and ${\bf B}_{R,N}$ are the nuclear fields in left in right dots that are static in a given run, ${\bf S}_L, {\bf S}_R$ are the left and right spin operators.  Orienting the external field in the $z$ direction, the $x$ and $y$ components of the nuclear field tend to admix $T_-, T_+$ and $S$.  Because of the large energy difference between these states in the presence of the large external magnetic field
$|{\bf B}_{\rm ext}|\gg |{\bf B}_{L,N}|,|{\bf B}_{R,N}|$, these transitions are suppressed.  However, the $z$ component of the nuclear field causes $S$ and $T_0$ to admix (out of the rotating frame) with a time scale $\tau_{\rm admix} =1/(B_{N} g \mu_B)$.  This time scale can in practice be larger or smaller than the inverse tunneling rate, $\Gamma^{-1}$.  

The magnetic field from the nuclear spins also causes the ESR pulse to be usually on resonance with only one of the spins.  We will now consider the two limiting cases, where tunneling is much faster or much slower than the $S, T_0$ admixing time.

\subsection{Slow triplet-singlet admixing}

In the first case where $T\gg \tau_{\rm admix} \gg \Gamma^{-1}$, the analysis
of the previous section is applicable, with the exception that the time  $T$
between successive cycles is much longer than $\tau_{\rm admix}$, so the $T_0$
component of the state will have time to admix with $S$ and subsequently
tunnel out (as was the case in the experiment.\cite{leo})  In this regime, we
have the case where the weak measurement only removes part of the singlet
portion of the state, but the projective measurement removes both the $S, T_0$
component.  Also, in the initial state of the cycle, there will be no $T_0$
component, so a different initial state is relevant.  Repeating the steps in
the minimal model section, starting with an initial state $(\beta, \gamma, 0,
0)$, taking the weak measurement on the singlet only but the projective
measurement over both $S$ and $T_0$, we find 
\bea
P_{\rm tot} &=& \sin^2\theta_1  \left(1-e^{-\Gamma t}\right) \vert \beta +\gamma\vert^2/2 \label{generalp2} \\
&+& \vert \beta + \gamma\vert^2 \left(\cos \theta_1 \sin \theta_2 + \sin \theta_1 \cos \theta_2 e^{-\Gamma t/2}\right)^2/2 \nonumber \\ &+& \vert \beta - \gamma\vert^2  \sin^2 (\theta_1 + \theta_2)/2.
 \nonumber
\label{case1}
\eea
Averaging this over a stochastic realization of the initial state preparation
yields $\la \vert \beta + \gamma\vert^2\ra_r = \la \vert \beta -
\gamma\vert^2\ra_r = 1/2$.  Notice that the first two terms are the same as
before, while the third term (arising from the $T_0$ projection) has no
exponential suppression from the weak measurement.

\subsection{Fast triplet-singlet admixing}

The more interesting case is that of fast $T_0, S$ admixing compared to the
tunneling time, $\tau_{\rm admix} \ll \Gamma^{-1}\ll T$.  In this case, the
$T_0$ and $S$ components quickly oscillate into one another, and the tunneling
process removes both the $S$ and $T_0$ component in a symmetric way.  This
physics can be implemented by applying an exponential decay POVM, Eq.~(\ref{s-POVM}), 
to both $S$ and $T_0$.  Repeating the measurement dynamics
analysis, it is straightforward to verify that the quantum system may be
effectively represented by a two dimensional quantum system (qubit), rather than
the four-level system above.  This qubit represents the two possibilities of
the single spins being parallel or anti-parallel with one another (this can
also be described as even or odd ``parity'').\cite{parity1,parity2,myparity1,myparity2}
Therefore, we can write an effective state $\psi = (\alpha, \beta)$, where
$\alpha$ represents the parallel amplitude, while $\beta$ represents the
anti-parallel amplitude.   

The manipulation steps described above now read as follows:  ({\it i})  The system always starts in the initial state $\psi = (1, 0)$, being in the spin blockade regime  (note that the initial state is the composite prepared state, not the ``microscopic'' (single-spin) one).  ({\it ii}) The ESR pulse on one spin then converts the spin-blockaded state into the prepared initial state, $\psi_{\rm ESR} = (\cos \theta_1, \sin \theta_1)$.  ({\it iii}) The POVM elements in the (parallel-antiparallel basis) now take the simple form
\be
M^\dagger_0 M_0 = \begin{pmatrix} 1 & 0 \\ 0 & e^{-\Gamma t}\end{pmatrix}, \quad 
M^\dagger_1 M_1 = \begin{pmatrix} 0 & 0 \\ 0 & 1-e^{-\Gamma t} \end{pmatrix},
\ee
implying
\bea
P_{iii}(1) &=& \sin^2 \theta_1 (1-e^{-\Gamma t}), \nonumber \\ 
P_{iii}(0) &=& \cos^2 \theta_1 + \sin^2 \theta_1 e^{-\Gamma t}. \label{yesno}
\eea
If the tunneling occurs, the state is destroyed, while if the tunneling does not occur, the post-measurement state is
\be
\psi' = \frac{1}{\sqrt{\cos^2 \theta_1 + \sin^2 \theta_1 e^{-\Gamma t}}} \begin{pmatrix} \cos \theta_1 \\ \sin \theta_1  e^{-\Gamma t/2} \end{pmatrix} \label{postm}.
\ee
Following Ref.~\onlinecite{Katz} for this simple case, we introduce the new angle ${\vartheta}$, so the state (\ref{postm}) may also be written as
\be
\psi' = \begin{pmatrix} \cos \vartheta \\ \sin \vartheta  \end{pmatrix} \label{postm2}.
\ee

({\it iv}) The second ESR pulse (again on one spin only) is now applied with rotation angle $\theta_2$, to produce the state 
\be \psi_{\rm final} = \begin{pmatrix} \cos (\vartheta +\theta_2)\\ \sin (\vartheta +\theta_2)  \end{pmatrix}. \label{finalstate}
\ee
({\it v}) Projecting on the anti-parallel state, we find $P_{v}(1) = \sin^2 (\vartheta +\theta_2)$, and $P_{v}(0) = \cos^2 (\vartheta +\theta_2)$.   This leaves the total probability of tunneling in the cycle as 
\bea 
P_{\rm tot} &=& \sin^2 \theta_1 (1-e^{-\Gamma t}) \\ &+&  \left(\cos^2 \theta_1 + \sin^2 \theta_1 e^{-\Gamma t} \right) \sin^2 (\vartheta +\theta_2).\nonumber 
\eea
In this expression, the preparation angle $\theta_1$ and the tomography angle $\theta_2$ 
provide a simple way of extracting the angle $\vartheta$ experimentally and verifying the theory of weak measurement in this system (as was similarly done in Ref.~\onlinecite{Katz}).

\subsection{Quantum Undemolition}

In this same regime, $\tau_{\rm admix} \ll \Gamma^{-1} \ll T$, it is also experimentally realistic to undo a measurement of an unknown initial state, ``quantum undemolition'' (QUD).\cite{QUD}  The idea follows that of the phase qubit introduced by Korotkov and one of the authors and relies on erasing the information obtained from the first measurement (for a popular version, see Ref.~\onlinecite{ns}).  
The first two steps follow the prescription above: First, prepare the initial state with any angle $\theta_1$, and make a weak measurement, characterized by the strength $\Gamma t$.  If no tunneling occurred, this brings us to the state (\ref{postm}).  Next, swap the parallel and antiparallel amplitude with a $\pi$-pulse on a {\it single spin}.  Next, make a second weak measurement of the same strength, $\Gamma t$, exactly as described above. Finally, a second $\pi$-pulse swaps the amplitudes again back to the initial state.  If the system did not tunnel in the first weak measurement, then the quantum state disturbance (\ref{postm}) occurred.
If the system did not tunnel in the second weak measurement, then the quantum state disturbance of the first measurement (\ref{postm}) is undone, fully restoring the initial state $\psi_{\rm ESR}$ (even if this state is unknown).  The probability for the QUD measurement to succeed, $P_S$, is simply the probability that the electron did not tunnel in the second measurement.  Given that the state disturbance (\ref{postm}) did occur, the QUD success probability is
\be
P_{S} = \exp ( -\Gamma t)/(\cos^2 \theta_1 + \sin^2 \theta_1 e^{-\Gamma t}).
\label{success}
\ee
This means that a successful QUD measurement becomes less likely as the measurement strength increases.\cite{note}

In order to confirm this theoretical prediction (for any initial state), it is necessary to make the further tomographic steps as in the minimal model section.  This is carried out with a tomographic ESR pulse (that can be combined with the last $\pi$-pulse) characterized by an angle $\theta_2$, and a projective measurement on the anti-parallel state.

  The total probability of transporting one charge is the additive probability of tunneling in one of the three attempts described above, where each attempt probability is the multiplicative probability of tunneling at that time, but not at any previous step. Following a similar analysis as before, we find that the total probability of tunneling at any step is
\be
P_{\rm tot} = 1-e^{-\Gamma t} + e^{-\Gamma t} \sin^2 ({\tilde \theta} +\theta_2).
\label{Pqud}
\ee
The angle $\tilde\theta$ is again to be extracted experimentally (similar to $\vartheta$ previously).
Here, we predict that ${\tilde \theta} =\theta_1$ if the measurement is undone.\cite{QUD}  The first term in (\ref{Pqud}) represents the possibility that tunneling occurs during the first weak measurement (no state disturbance to begin with), or the second weak measurement (a failed undoing attempt).  The last term in (\ref{Pqud}) describes a successful QUD measurement, where the post-measurement state of the undoing measurement coincides with the initial prepared state (regardless of our knowledge of it).  Notice that the prefactor of the last term in (\ref{Pqud}),  $e^{-\Gamma t}$, is interpreted as the QUD success probability (\ref{success}) times $P_{iii}(0)$, the probability the initial state disturbance occurred in the first place (\ref{yesno}). 

Note that (\ref{Pqud}) recovers the correct limits:  If $t=0$, the two $\pi$-pulses simply undo each other, and the two angles add.  As $t \rightarrow \infty$, there is always a transported charge: the first measurement removes the anti-parallel component; the $\pi$-pulse and the second measurement removes the parallel component.

Another interesting property of the undemolition sequence described above
is that it can undo unitary errors that occur due to the presence of the 
nuclear spins, as described by Eq.~(\ref{H-nucl}) where the nuclear spins are
treated classically and in the case when the
external magnetic field exceeds the nuclear field.  In that case, the nuclear field
leads to a random phase $\varphi$ between the parallel (upper) and antiparallel (lower) 
component of the state Eq.~(\ref{postm}).  The term $i\varphi$ is then simply
added to $\Gamma t$ in the exponent in the antiparallel component, and is erased
in much the same way as the $\Gamma t$ contribution due to the weak measurement.\cite{note2}
This spin echo-like effect has the added advantage that after many realizations, the other terms in Eq.~(\ref{Pqud})
will be suppressed by averaging over the uncontrolled phase $\varphi$ that will change from run to run, while the important undemolition term will remain, protected from the influence of the uncontrolled nuclear spins.

\section{Dephasing and spin resonance on both spins}

We now discuss in more detail the effect of dephasing on our results. 
We assume that the measurement time scales are much shorter than the ESR time
scales, and therefore only include the effect of the Hamiltonian dynamics with
the ESR manipulation (though the combination of both unitary and nonunitary
dynamics is also very interesting, see Ref.~\onlinecite{korotkovboth}).  These
unitary operations may be included into the analysis by operating with a
generalized version of (\ref{Rl}),  
\be
R_\alpha = \begin{pmatrix}  \cos \theta_\alpha  e^{i\phi_\alpha} & -\sin \theta_\alpha \\ \sin \theta_\alpha & \cos \theta_\alpha e^{-i\phi_\alpha}\end{pmatrix},
\label{Rl2}
\ee
where $\alpha=L,R$, the angle $\theta$ is the rotation angle about the $y$-axis, and $\phi$ is the rotation angle about the $z$-axis.  In the two-spin (left/right) Hilbert space, the above unitary operation on the left spin is given by
\be
U_L = \begin{pmatrix} \cos \theta_L e^{i\phi_L} & 0 & 0 & -\sin \theta_L \\ 
0 & \cos \theta_L e^{-i\phi_L} & \sin \theta_L & 0\\
0 & -\sin \theta_L & \cos \theta_L e^{i\phi_L} & 0 \\
\sin \theta_L & 0& 0 & \cos \theta_L e^{-i\phi_L} \end{pmatrix}.
\ee
The same on the right spin is given by
\be
U_R= \begin{pmatrix} \cos \theta_R e^{i\phi_R} & 0 & -\sin \theta_R & 0 \\ 
0 & \cos \theta_R e^{-i\phi_R} & 0 &  \sin \theta_R\\
 \sin \theta_R& 0 & \cos \theta_R e^{-i\phi_R} & 0 \\
0 & -\sin \theta_R& 0 & \cos \theta_R e^{i\phi_R} \end{pmatrix}.
\ee
The commuting matrices may be applied together with the ESR manipulations, and subsequently averaged over a Gaussian random distribution, whose width is controlled by the strength of the magnetic field fluctuations.  A full analysis of dephasing is quite involved because statistically independent phases enter at every step in the procedure.  Here, we present a simpler analysis that captures the basic physics.  We consider the most important process of a dynamically changing $z$-component of the nuclear magnetic field that affect both spins in the same way.  The process is modeled by introducing a phase $\phi_L =\phi_R = \phi = (g \mu_B/2\hbar) \int_0^t B_N(t') dt'$ on both left and right spins, with $(\theta_1, \phi_1)_L$ and $(0, \phi_1)_R$ for the first ESR pulse, and $(\theta_2, \phi_2)_L$ and $(0, \phi_2)_R$ for the second ESR pulse.  We will then average over the phase $\phi(t)$ assuming uncorrelated white noise,\cite{noise} 
$\la B_N(t') B_N(t'')\ra = \sigma^2 \delta(t'-t'')$, so that
\bea
\la \phi(t) \phi(0)\ra &=& (g \mu_B/2\hbar)^2 \int_0^t dt' dt'' \la B_N(t') B_N(t'')\ra \nonumber \\
&=& (g \mu_B/2\hbar)^2 \sigma^2 t =  D t,\label{dephase}
\eea
where we introduced the dephasing rate, $D$.

Repeating the treatment in the minimal model section, starting with a general triplet state $(\beta, \gamma, \alpha, 0)$ in the $T/S$ basis, we find that the total probability before any averaging is
\bea
P_{\rm tot} &=& (1-e^{-\Gamma t})\sin^2 \theta_1 \vert \beta e^{i \phi_1} + \gamma e^{-i \phi_1}\vert^2/2  \\
&+& \vert -\alpha \sin \theta_1 \sin \theta_2 (e^{i \phi_1 + i \phi_2}-e^{-i \phi_1 - i \phi_2})/\sqrt{2} \nonumber \\
&+& \sin \theta_1 \cos \theta_1 e^{-\Gamma t/2} (\gamma e^{-i \phi_1} + \beta e^{i \phi_1}) \nonumber \\
&+& \cos\theta_1 \sin \theta_2 (\beta e^{2 i \phi_1 + i \phi_2} + \gamma e^{-2 i \phi_1 - i\phi_2}) \vert^2/2. \nonumber
\eea
If we now average over both initial state preparation, $\la|\alpha|^2\ra_{r} = \la|\beta|^2\ra_r= \la |\gamma|^2\ra_r=1/3$, (with vanishing averaged initial coherence), as well as the nuclear field, we end up with
\bea
\la P_{\rm tot}\ra &=& [\sin^2 \theta_1 (1-e^{-\Gamma t}) +\sin^2 \theta_1\sin^2 \theta_2 e^{-D \tau}\sinh(D\tau) \nonumber \\
&+& \sin^2 \theta_1\cos^2 \theta_2 e^{-\Gamma t} + \cos^2 \theta_1 \sin^2 \theta_2 \nonumber \\
&+& 2 \sin \theta_1 \cos \theta_1 \sin \theta_2 \cos \theta_2 e^{-D \tau - \Gamma t/2}]/3,
\eea
where $\tau =\tau_1 +\tau_2$ is different from the weak measurement time $t$.
Here we see the presence of the $T_0$ term that coherently canceled before, as
well as the suppression of the interference term that scales as $e^{- \Gamma
  t/2}$.  Other types of dephasing will act similarly, suppressing all the
terms in general. 

\section{Conclusions}

We have developed a theory of weak quantum measurements for spin
qubits. Inspired by a recent experiment demonstrating single-spin manipulation
with ESR pulses in a double quantum dot setup, we have shown how the current
through such a device is affected by the fact that a quantum measurement of
the spin state can be either weak or projective. The system is
operated in the spin blockade regime, where the spin singlet state contributes
to transport and the three spin triplet states block it. A sequence of a state
preparation step (using ESR), a weak measurement step (using electron
tunneling and spin-to-charge conversion), a state tomography step (using ESR),
and a final strong projective measurement step (using electron
tunneling and spin-to-charge conversion) is sufficient to exhibit a clear
signature of quantum weak measurement in the current through the
double quantum dot system. We have analyzed how our results are affected by
spin dephasing. As the major source of dephasing we have discussed the hyperfine
interaction of the electron spin with the surrounding nuclear spins of the
substrate. This is a well established fact for GaAs quantum dots. We have shown
that the combined effects of singlet-triplet mixing due to the nuclear field
plus the consequences of weak measurement theory yield interesting
results in the regime where the singlet-triplet mixing is fast compared to the
tunneling time of the weak measurement step. In this regime, the weak
measurement can even be undone and spin-echo technique are applicable in a
straightforward way. We believe that our predictions can be readily observed
in spin qubits formed, for instance, in GaAs quantum dots. 

\section{Acknowledgments}

We would like to thank Frank Koppens and Lieven Vandersypen for interesting
and inspiring discussions.
ANJ thanks Christoph Bruder and the Basel theory group for kind hospitality.
This work was financially supported by the Swiss NSF and the NCCR Nanoscience.

\bibliographystyle{apsrev}

\begin{thebibliography}{99}
\bibitem{tomo}
Here we use only one tomography angle because the experiment, Ref.~\onlinecite{leo}, can make rotations about one axis.
To do full quantum state tomography, controlled ESR rotations about both the $x$ and $y$-axis are necessary, or the tunnel coupling between the dots must also be pulsed.
\bibitem{note}
It is straightforward to show this undoing is optimal in the sense of Ref.~\onlinecite{QUD}.
\bibitem{noise}
The experiment, Ref.~\onlinecite{leo}, found the nuclear spin environment was actually non-Markovian, but this analysis is outside the scope of the present article.
\end{thebibliography}

\end{document}